# Nanomagnetic Planar Magnetic Resonance Microscopy "Lens"

Mladen Barbic
Department of Physics and Astronomy, California State University, Long Beach
1250 Bellflower Boulevard, Long Beach, California 90840 USA

and

Axel Scherer
Department of Electrical Engineering, California Institute of Technology
1200 E. California Boulevard M/C 200-36, Pasadena, California 91125 USA

**Abstract**

The achievement of three-dimensional atomic resolution magnetic resonance microscopy remains one of the main challenges in the visualization of biological molecules. The prospects for single spin microscopy have come tantalizingly close due to the recent developments in sensitive instrumentation. Despite the single spin detection capability in systems of spatially well-isolated spins, the challenge that remains is the creation of conditions in space where only a single spin is resonant and detected in the presence of other spins in its natural dense spin environment. We present a nanomagnetic planar design where a localized Angstrom-scale point in three-dimensional space is created above the nanostructure with a non-zero minimum of the magnetic field magnitude. The design thereby represents a magnetic resonance microscopy "lens" where potentially only a single spin located in the "focus" spot of the structure is resonant. Despite the presence of other spins in the Angstrom-scale vicinity of the resonant spin, the high gradient magnetic field of the "lens" renders those spins inactive in the detection process.

Electronic mail: mbarbic@csulb.edu

The original reports in 1973 by Lauterbur [1] and Mansfield and Grannell [2,3], have introduced Magnetic Resonance Imaging (MRI) as an invaluable three-dimensional visualization technology with a great impact in clinical medicine. Although improvements in imaging resolution through conventional inductive detection [4,5] have steadily progressed within the last three decades, present spatial resolution is limited to approximately 1μm in nuclear and electron spin magnetic resonance microscopy [6-9]. Despite the challenges, the promise of a 3D atomic resolution MRI with the well-known advantages of a non-invasive, three-dimensional, multi-contrast, and chemically specific imaging tool [10,11] remains very attractive. The challenge in improving the imaging resolution results from the extremely weak signals in the magnetic resonance process [12], spin diffusion, and the limited ability to create sufficiently large gradient fields by current carrying coils.

Motivated by the potential of combining 3D imaging capability of conventional magnetic resonance with the atomic resolution of scanning probe techniques that utilize mechanical cantilevers, Sidles proposed a magnetic resonance imaging technique with potential for atomic resolution single spin detection [13]. This method, magnetic resonance force microscopy (MRFM), uses a microscopic magnetic particle as a source of atomic scale imaging gradient fields and a mechanical resonator as a sensitive detector of magnetic resonance [14]. Proof-of-concept demonstrations of the technique were reported for both electron [15] and nuclear spin resonance [16]. The progress in MRFM has recently culminated in the mechanical detection of a single electron spin [17].

Further progress of this microscopy technique, however, places challenging demands on sensitivity and resolution requirements. Mechanical detection of a single electron spin magnetic resonance can be performed on a sample with spatially well-isolated spins, but only after significant averaging time of 13 hours per data point [17]. In order to improve the sensitivity and reduce the averaging time, novel fabrication methods for the miniaturization of all the critical components in MRFM (the mechanical detector, magnetic field gradient source, and optical nanoreflector) are being developed [18]. Further progress in experimental sensitivity improvements is certain to continue until single nuclear spin detection is accomplished, a feat that would be significant in molecular imaging applications.

In this article, however, we focus on the magnetic resonance imaging resolution problem. Even if single spin detection capability becomes readily available through signal-to-noise improvements, it is unlikely that in the natural dense spin environments (that one would ultimately want to image with atomic resolution), only a single spin would be resonant while the neighboring spins would not contribute to the detected signal. This statement is deduced from the slice-selective nature of MRFM imaging technique where all of the spins for which the resonant condition:

$$\omega(r) = \gamma |B(r)| \qquad (1)$$

is satisfied generally contribute to the detected signal. It is important to emphasize that equation (1) is a scalar relationship between the resonant frequency of the spin, $\omega$, and the magnitude of the magnetic field, $|B|$, at the location of the spin, where $\gamma$ is the gyromagnetic ratio for the nuclear or electron spin. It is the magnitude of the magnetic field at the spin location that determines its resonant frequency, and therefore the slices of constant |B| have to be well understood in order to deconvolve and reconstruct [19-23] the image from the available data. Generally, due to the size of the large polarizing field that must be applied to the sample, only the z-component of the magnetic field from the gradient sources is considered [10,11]. However, that is an approximation only, and needs to be carefully reconsidered when lower magnetic fields are utilized.

The question that we address is whether it is possible, for the purpose of atomic resolution MRI, to create a point in three-dimensional space where the magnitude of the magnetic field is a non-zero extremum, so that equation (1) is satisfied not for a slice, as has been considered so far in MRFM, but for a point in space. If that is possible on the atomic scale, then it is conceivable that only a single spin in three-dimensional space could be resonant and detected in the presence of other non-interfering nearby spins. Drawing on the previous advances in other scientific disciplines, we demonstrate a nanomagnetic planar design representing a magnetic resonance microscopy "lens" with a non-zero minimum of the magnetic field magnitude "focus" located away from the plane of the structure. Despite the presence of other spins in the vicinity of the resonant spin, the high gradient magnetic fields of the "lens" ensure that those spins remain off-resonance and thereby undetected. The "lens" structure presented here for the potential atomic resolution magnetic resonance imaging might provide magnetic field properties desirable in other scientific fields such as the diamagnetic or neutral particle trapping and levitation, and quantum computation.

Maxwell's equations place restrictions on the properties of magnetostatic fields in free space. It is impossible for the magnitudes of the components of the magnetic field vector $B_X$, $B_Y$, or $B_Z$ to have a local minimum or maximum in free space [24].

Additionally, the magnetic field magnitude, $|\mathbf{B}|$, cannot have a local maximum, but *can have* a local minimum in free space [25]. These properties have received much attention in the fields of plasma confinement [26,27], neutral particle trapping [25,28-31], and diamagnetic levitation [32-34], but, to our knowledge, have not been considered in magnetic resonance imaging. Of particular interest to us is a configuration that produces a local non-zero magnetic field magnitude $|\mathbf{B}|$ minimum in free space [28-30], since non-zero magnetic field is required to obtain magnetic resonance per equation (1).

Figure 1(a) shows our nanomagnetic planar magnetic resonance microscopy "lens" design. It consists of a thin circular disk of magnetic material in the x-y plane with a perpendicular anisotropy axis so that it is permanently magnetized along the z-direction (out of the page). In addition, two quarter-circle cuts are made in the disk, diagonally opposed and with a smaller radius. Two axes of symmetry along +45 and –45 degrees are indicated. In order for the "lens" to have a local out-of-plane non-zero magnetic field magnitude minimum, a constant bias magnetic field in the direction opposite to the magnetization direction of the structure (into the page) is also required. In this article, we use the following parameters for our design: (a) the perpendicular anisotropy magnetic material has $\mu_0 M = 2$(Tesla), (b) the outer radius of the structure is 60(nm), (c) the inner radius of the structure is 40(nm), (d) the thickness of the structure is 10(nm), and (e) the bias magnetic field in direction opposite to the magnetization is $B_{BIAS} = -650$ (Gauss). We believe that the magnetic material is well within reach of perpendicular anisotropy magnetic thin film technology currently used in the magnetic recording industry, and the available coercivity is more than sufficient to sustain the opposing magnetic field of –650 (Gauss) [35]. The dimensions of the nanostructure are well within the capabilities of the state-of-the-art lithography and focused-ion-beam (FIB) technology [36]. The selection of the magnetic material used in the design will also have to include careful consideration of magnetic fluctuations [37,38] in order to minimize quantum decoherence [39].

Our design is a planar permanent magnet analogue of the planar Ioffe trap design of Weinstein and Libbrecht that utilizes current carrying wires (the Ioffe trap (c) in Reference 29). This is best described by showing, in Figure 1(b), the effective circulating Amperian pseudo-currents [40] within our structure. From this perspective, the structure consists of (a) one outside full-loop current, (b) two quarter-loop inner currents running in opposite direction to the outside current loop, and (c) four Ioffe bars. In addition, all of the currents have the same magnitude, and the outside radius is 1.5 times the inner radius of the structure, completing the analogy between the nanomagnetic planar magnetic resonance microscopy "lens" and the Weinstein-Libbrecht current carrying planar Ioffe trap [30]. It is important, however, to distinguish the advantages of a permanent magnet design for magnetic resonance imaging or neutral particle trapping. As stated by Halbach [41]: *"when it is necessary that a magnetically significant dimension of a magnet is very small, a permanent magnet will always produce higher fields than an electromagnet"*, and *"can be scaled to any size without any loss in field strength"*. Miniaturization of a permanent magnet also provides an increase in magnetic field gradients and curvatures required for ultra-high magnetic resonance imaging resolution. Additionally, the presented planar permanent magnet design requires no outside power supply and no interconnecting leads. Finally, due to the quantum mechanical exchange interaction responsible for ferromagnetism of the structure, the system generates no heat and requires no heat dissipation.

We model the magnetic fields above the structure by assuming a uniform magnetization **M** directed along the z-axis. Therefore the uniform positive magnetic pseudo-surface-charge density of **n·M** is on the top surface, where **n** is the outward normal of the magnet, and the corresponding negative pseudo-surface-charge density is on the bottom surface. We numerically compute first the scalar potential φ(**r**) and then the magnetic field **B**(**r**) at a position **r** above the plane of the "lens":

$$\vec{B}(\vec{r}) = -\mu_0 \nabla \varphi(\vec{r}) = -\mu_0 \nabla \left( \sum_{i=1}^{2} \frac{1}{4\pi} \int_{S_i} \frac{\hat{n}_i \cdot \vec{M}_i}{|\vec{r} - \vec{r}'|} da' \right) \quad (2)$$

where $\mu_0$ is the permeability of free space. Additional bias field in the negative z-direction $B_{BIAS}$= -650(Gauss) is also applied. Figures 2(a) and 2(b) show the contours of constant magnitude of magnetic field **B** above the structure along the two symmetry planes at +45 and –45 degrees. The figures are 20nm by 20nm in size. As intended, we find that the structure produces a non-spherical magnetic field magnitude minimum above the plane. For the parameters used, the minimum is located 23.8(nm) above the surface and has a value of $B_{MIN}$=99.5(Gauss). The contours are 6 (Gauss) apart, with the center contour at 100.5 (Gauss). The localized minimum of the magnetic field magnitude only occurs if bias magnetic fields between $B_{BIAS}$=-550(Gauss) and $B_{BIAS}$=-750(Gauss) are applied, otherwise a saddle structure is observed in the magnetic field magnitude profiles.

Putting aside the sensitivity issue for the moment, the imaging resolution of the nanomagnetic planar magnetic resonance microscopy "lens" depends on the magnetic field curvatures it produces and the characteristic linewidth of the spin resonance. A spin resonance linewidth of ~1(Gauss), typical of nuclear spins in a solid state environment, would mean that the "lens" would be able to frequency separate different spins located approximately 1(nm) to 4(nm) apart, as only spins located within the central contours in the Figures 2(a) and 2(b) would be resonant. Despite the potential presence of the nearby spins in the nanometer-scale vicinity of the spins in "focus", high gradient magnetic field of the "lens" would render those spins non-resonant, as the condition of equation (1) would not be satisfied for these neighboring spins. It should be noted that the resonance linewidth of spins in isolated molecules on surfaces could be much narrower, and potential radio-frequency pulse sequences applied to samples might reduce the resonance linewidths by several orders of magnitude, resulting in the Angstrom scale "focus" of the "lens". The consequence of this capability would mean that the three-dimensional atomic resolution magnetic resonance microscopy is possible, as far as the gradient magnetic fields are concerned. For the sub-surface 3-D single electron or proton spin imaging, the angstrom scale motion of the "lens" can be achieved using the well-developed piezoelectric scanning probe microscopy technology.

At this stage, the sensitivity limits in magnetic resonance microscopy are being intensively pursued, and the sensitivity required for single nuclear spin detection remains to be demonstrated, apart from the already demonstrated single electron spin detection [17]. In addition to more advanced micro/nano-mechanical force detectors [18], several other sensing mechanisms remain viable candidates for improving the imaging sensitivities in detection of small number of spins. They include the measurement of direct transfer of angular momentum [42,43] and energy [44,45] to the spin population in magnetic resonance using micro-mechanical cantilevers, flux-detection class of magnetic resonance detection schemes such as micro-coil NMR [46,47], superconducting quantum interference devices (SQUID) [48,49], Hall sensors [50], superconducting resonators [51], and optical methods [52,53]. Additionally, a single or few spin detection schemes will likely require new methodologies in the area of quantum measurement [54-57] that deviate significantly for the classical theory of magnetic resonance detection, and have to involve careful consideration of spin polarization [14] and spin noise [58] in a few-spins regime.

Before concluding, we discuss one particular point of significance in force detection using the magnetic resonance microscopy "lens" fabricated on a sufficiently sensitive cantilever. Although the magnitude of the magnetic field is designed to have a localized minimum in three-dimensional space above the plane of the structure, the force on the resonance spin can be non-zero. This can best be seen by showing the vector plots of the magnetic fields around the magnetic field magnitude minimum along the two symmetry planes of the structure in Figures 3(a) and 3(b). The size of the view is 10nm by 10nm centered at the location of the magnetic field magnitude minimum at z=23.8(nm). Variation of the magnetic field vector direction and magnitude is clearly visible from the plots. Although the magnetic field vector length has a localized minimum, the components of the magnetic field vector vary along different directions, and that field gradient could be used in the force detection. The force on the spin is:

$$\vec{F} = \nabla(\vec{m} \cdot \vec{B}) \qquad (3)$$

and therefore the components of the force on the spin depend on the tensor of the gradient of the magnetic field. Inspection of the vector plots and numerical analysis reveal that only the two gradient components of the gradient magnetic field tensor are non-zero at the "focus":

$$\begin{pmatrix} \frac{\partial B_X}{\partial x} & \frac{\partial B_Y}{\partial x} & \frac{\partial B_Z}{\partial x} \\ \frac{\partial B_X}{\partial y} & \frac{\partial B_Y}{\partial y} & \frac{\partial B_Z}{\partial y} \\ \frac{\partial B_X}{\partial z} & \frac{\partial B_Y}{\partial z} & \frac{\partial B_Z}{\partial z} \end{pmatrix} = \begin{pmatrix} -2.5 & 0 & 0 \\ 0 & 2.5 & 0 \\ 0 & 0 & 0 \end{pmatrix} \qquad (4)$$

Therefore, even if the magnetic field magnitude has a localized extremum at the "focus" of the "lens", the resonant spin can experience a non-zero gradient field of 2.5(Gauss/Angstrom) and be force-detected if sufficient sensitivity and appropriate quantum measurement methodology is available. A "focus" field of ~100 Gauss for our "lens" design corresponds to the ~425kHz proton resonance frequency, a convenient frequency, as it can closely couple to the common resonant frequencies of micro/nano-mechanical resonators. We also point out, that despite our main focus on atomic resolution microscopy, the structure we presented in this article is completely scalable. Following Halbach's argument [41], a larger structure will produce the same magnetic field features, albeit at a reduced spatial resolution performance, but at a level still sufficient for ultra-high resolution magnetic resonance microscopy.

We conclude by noting that, since the development of scanning probe microscopes [59, 60], atomic resolution imaging has been limited to two-dimensional surfaces. It is worth pointing out that these techniques depend on the conduction band or Coulomb interaction between the outermost atomic electrons on the probe and the outermost atomic electrons on the surfaces. Proton magnetic resonance imaging, on the other hand, is intrinsically a spectroscopic hydrogen nucleus detection technique that utilizes spatially varying magnetic fields to achieve non-invasive three-dimensional imaging. It is our hope that the concept of a nanomagnetic planar magnetic resonance microscopy "lens" (or variations of the presented idea) can extend the scanning probe and magnetic resonance imaging capability into the three-dimensional atomic resolution regime.

This work has been generously supported by the Caltech Grubstake Fund and by the National Science Foundation NSF-CAREER Award (DMR-0349319). The authors thank Dr. Joyce Wong and George Maltezos for helpful discussions, suggestions, and careful reading of the manuscript, and acknowledge initial discussions of potential applications of the presented nanomagnetic design towards neutral particle trapping and quantum computation with Professor Hideo Mabuchi and Ben Lev of Caltech.

Figure Captions

**Fig. 1.** (a) Nanomagnetic planar magnetic resonance microscopy "lens" design consists of a 10nm thick perpendicular anisotropy magnetic material disk magnetized out-of-page with two inside quarter-circle diagonally opposed cuts. Outside radius is 60nm and inside radius is 40nm, with two symmetry planes axes at +45 and –45 degrees as indicated. A bias field opposite to the magnetization direction is also required for obtaining a localized magnetic field magnitude minimum. Fig. 1. (b) The effective circulating Amperian pseudo-currents of the "lens" structure reveal one outer counter-clockwise full-loop current, two inner clock-wise quarter-loop currents, and four planar Ioffe bars.

**Fig. 2.** Contours of constant magnitude of the magnetic field above the "lens" structure in (a) along the +45 degrees symmetry plane and in (b) along the –45 degrees symmetry plane. The "focus" or the localized magnetic field magnitude minimum is located 23.8(nm) above the plane and has a value of 99.5 (Gauss). The contours are drawn at 6 (Gauss) intervals with the central contour at 100.5 (Gauss). Only the spins within the central contour satisfy the magnetic resonance condition, and would potentially be detected by a narrow bandwidth resonant detector, such as a nano-mechanical cantilever with a quality factor of Q~10,000-100,000. 20nm by 20nm areas are shown.

**Fig. 3.** Vector plots of the magnetic field vector above the plane of the "lens" structure along the two symmetry planes at (a) +45 degrees and (b) –45 degrees. 10nm by 10nm areas are shown. The variation of the magnetic field vector magnitude and direction is visible, and the variation of the magnetic field components through the central minimum provides the force gradients for the potential mechanical detection of magnetic resonance.

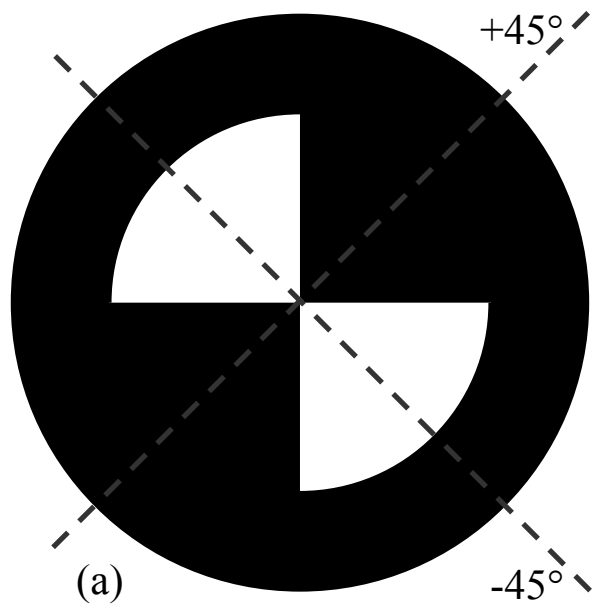

(a)

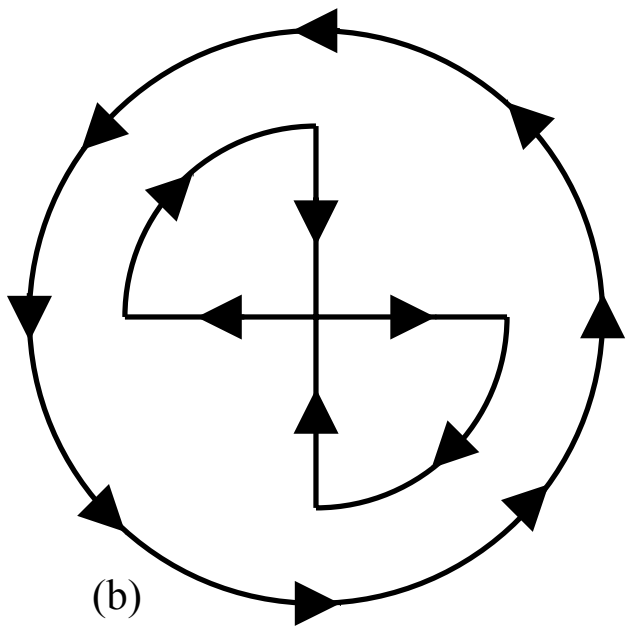

(b)

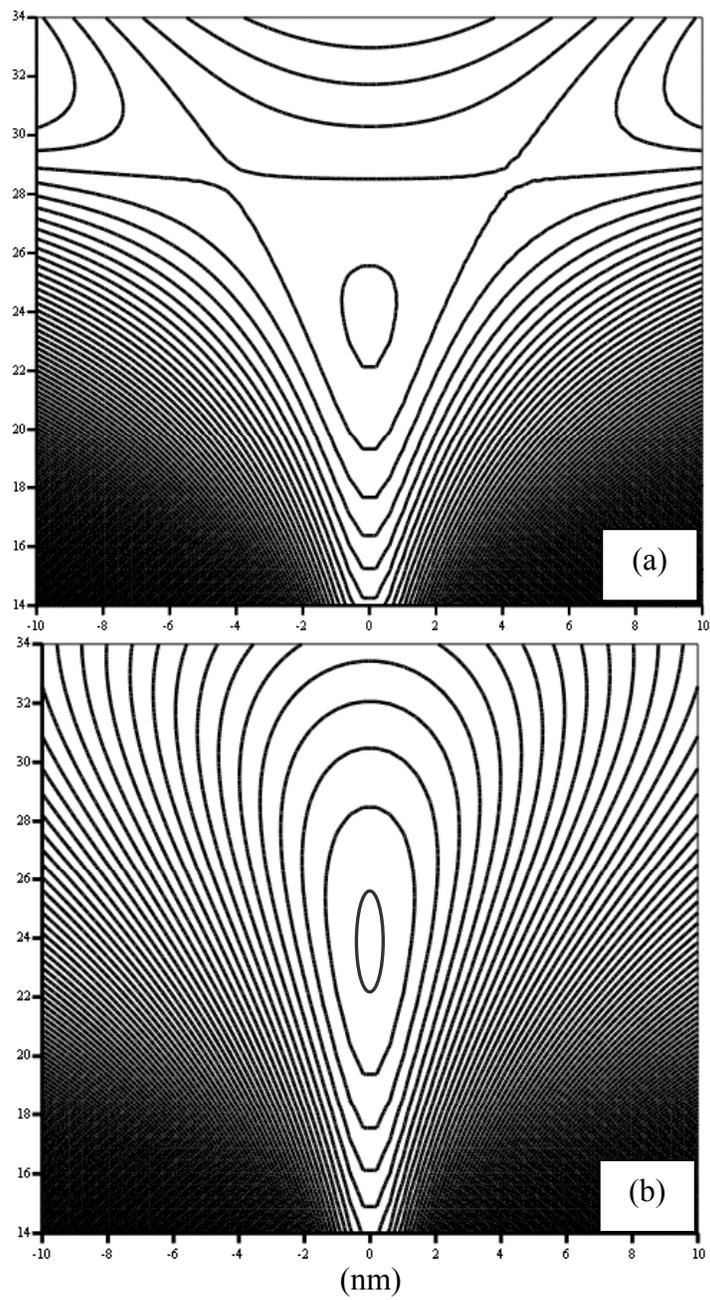

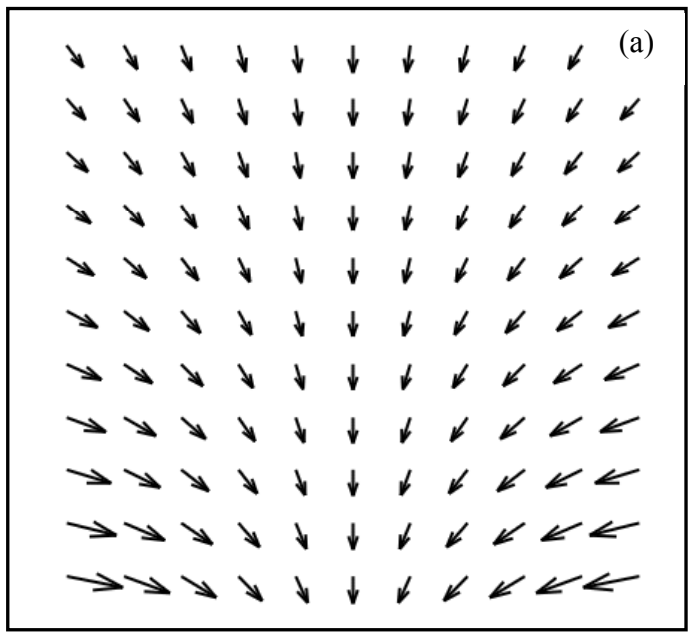

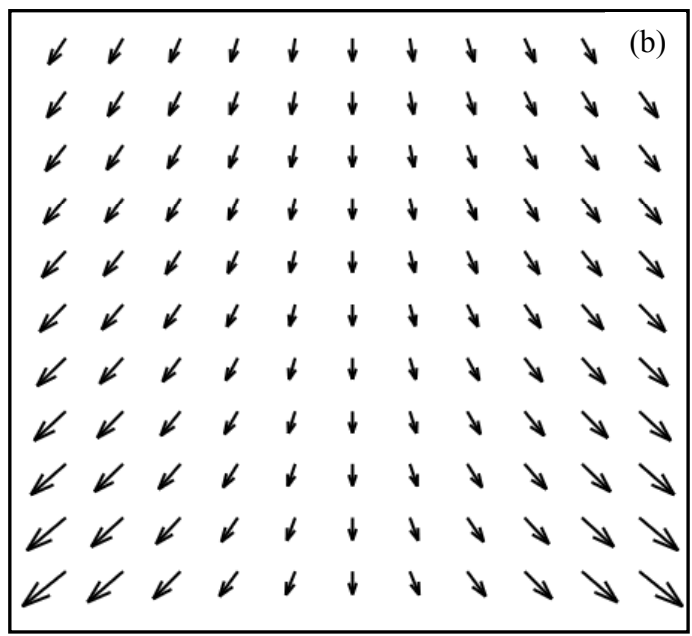

10(nm)